\documentstyle[twocolumn,twoside]{article}
\title{\Large \bf Five-Dimensional Tangent Vectors in Space-Time \\
\large \bf I. Introduction and Formal Theory}
\author{Alexander Krasulin \\ \it Institute for Nuclear Research of the
Russian Academy of Sciences \\ \it 60th October Anniversary Prospect, 7a,
117312 Moscow, Russia}
\date{\normalsize \bf Abstract \\ \mbox{ } \\ \begin{minipage}{400pt}
\normalsize
In this series of papers I examine a special kind of geometric objects
that can be defined in space-time --- five-dimensional tangent vectors.
Similar objects exist in any other differentiable manifold, and their
dimension is one unit greater than that of the manifold. Like ordinary
tangent vectors, the considered five-dimensional vectors and the tensors
constructed out of them can be used for describing certain local quantities
and in this capacity find direct application in physics. For example, such
familiar physical quantities as the stress-energy and angular momentum
tensors prove to be parts of a single five-tensor. In this part of the
series five-dimensional tangent vectors are introduced as abstract
objects related in a certain way to ordinary four-dimensional tangent
vectors. I then make a formal study of their basic algebraic properties
and of their differential properties in flat space-time. In conclusion
I consider some examples of quantities described by five-vectors and
five-tensors. \end{minipage} }
\begin{document}

\maketitle

\begin{flushleft}
\bf 1. Introduction
\end{flushleft}
Adding a dimension to tangent vectors in space-time is not a new idea in
physics. A well-known example is the Kaluza--Klein model [1] and the models
that succeeded it, where the extra dimension of tangent vectors results from
adding a dimension to the space-time manifold itself. Another example are the
theories of gravity formulated as Yang--Mills gauge theories of the de Sitter
group [2] and similar models, where the additional dimension is assigned not
to the tangent vectors themselves, but to the internal vector space where
the vierbein field takes its values. Unlike all these constructions, for
introducing the five-dimensional vectors I consider in this paper one does
not need to change the space-time manifold in any way nor to endow it with
any additional structure. The vectors I am going to discuss here, which I
will call {\em five-dimensional tangent vectors} or simply {\em five-vectors},
should be viewed as another type of geometric objects that can be defined
in space-time and which are more suited for describing certain kinds of
geometric and physical quantities than ordinary tangent vectors and tensors.

A hint to the existence of five-dimensional tangent vectors can be found in
spinors. For the type of 4-spinors commonly used in physics, the symmetry
group of the corresponding Clifford algebra is SO(3,2). Accordingly, there
exist five constituents of the Clifford algebra (five matrices) $\Gamma_{A}$,
where $A$ runs 0, 1, 2, 3, and 5, that all transform alike under Dirac and
charge conjugation:
\begin{equation}
\bar{\Gamma}_{A} = \Gamma_{A} \; \mbox{ and } \;
\Gamma_{A}^{\; c} = \Gamma_{A},
\end{equation}
and that satisfy the following anticommutation relations:
\begin{equation}
\Gamma_{A} \Gamma_{B} + \Gamma_{B} \Gamma_{A} =  - \, 2 \, \eta_{A B},
\end{equation}
where $\eta_{A B} \equiv {\rm diag}(+1,-1,-1,-1,+1)$. It is evident that one
can obtain a new set of five constituents satisfying the same conjugation
and anticommutation relations by applying an arbitrary O(3,2) transformation
to the original set. Moreover, any two sets of constituents that satisfy
relations (1) and (2) prove to be connected by an O(3,2) transformation. For
an appropriate choice of the constituent set, the standard $\gamma$-matrices
(the ones identified with the components of the basis four-vectors) are
expressed in terms of $\Gamma_{A}$'s as
\begin{equation}
\gamma_{\mu} = \frac{i}{2} (\Gamma_{\mu}\Gamma_{5} - \Gamma_{5}\Gamma_{\mu}),
\end{equation}
where $\mu$ = 0, 1, 2, or 3.

These observations may give one the idea to consider a new type of vectors
that make up a real five-dimensional vector space, $V_{5}$, endowed with
a symmetric nondegenerate inner product with the signature $(+---+)$ or
$(-+++-)$. Considering the relation that exists between multiplication in a
Clifford algebra and exterior multiplication of multivectors and forms, on
the grounds of equation (3) one may further suppose that there should exist
a certain correspondence between four-dimensional tangent vectors and part
of the bivectors constructed from elements of $V_{5}$, such that for each
orthonormal five-vector basis ${\bf e}_{A}$ from a certain class of such
bases, the four-vectors corresponding to the four bivectors ${\bf e}_{\mu}
\wedge {\bf e}_{5}$ make up a Lorentz basis in the space $V_{4}$ of ordinary
tangent vectors.

Basing on these assumptions one can make a formal study of the basic
algebraic and differential properties of five-dimensional tangent vectors, as
it is done in this paper. This formal analysis may serve as an introduction
to the subject and as a guide in developing a more sophisticated theory
of five-vectors basing on the principles of differential geometry, which
is presented in part II. Within this latter theory five-vectors are
introduced first as equivalence classes of parametrized curves and then, more
rigorously, as a particular kind of differential-algebraic operators that
act upon scalar functions. In part III I consider some applications of
five-vectors in flat space-time and of their analogs in three-dimensional
Euclidean space. In particular, I show how five-vectors can be used for
describing in a coordinate-independent way finite and infinitesimal Poincare
transformations and, as an illustration, reformulate the classical mechanics
of a perfectly rigit body in terms of the analogs of five-vectors in
three-dimensional space. In that part I also introduce the notion
of a ``bivector'' derivative for scalar, four-vector and four-tensor fields
in flat space-time and, as an illustration, calculate its analog in
three-dimensional Euclidean space for the Lagrange function of a system of
several point particles in classical nonrelativistic mechanics.

The fact that five-dimensional tangent vectors and the tensors associated
with them enable one to give a coordinate-independent description to finite
and infinitesimal Poincare transformations and, as one will see below, to
describe as a single local object such quantities as the stress-energy
and angular momentum tensors, should be thought of only as a reason for
considering five-vectors in the first place and for making an exploratory
study of their basic properties. If this were all there is to it, i.e.\ if
five-vectors only enabled one to present certain geometric quantities and the
relations between them in a mathematically more attractive form, such vectors
would hardly be of particular interest both to physicists, who typically do
not care much for fancy mathematics unless it enables them to formulate new
physical concepts, and to mathematicians, who would consider five-vectors as
merely a particular combination of already known mathematical constructions.
A more important reason why the concept of a five-dimensional tangent vector
is worth considering is that it enables one to extend the notion of the
affine connection on a manifold and of the connections which physicists call
gauge fields, and thereby at no cost at all, i.e.\ without changing the
manifold in any way and without introducing new gauge groups, to obtain
new geometric properties of space-time in the form of a new kind of torsion
and a new kind of gauge fields.

Before discussing these applications of five-vectors, in part IV I develop
a five-vector generalization of exterior differential calculus, which is more
a technical necessity---a necessity in replacing ordinary tangent vectors
with five-vectors in all the formulae related to integration of differential
forms and to exterior differentiation of the latter. Apart from allowing one
to present certain relations in a more elegant form, for scalar-valued forms
this generalization is equivalent to ordinary exterior calculus, which
was to be expected since five-vectors in this case are used only for
characterizing the infinitesimal elements of integration volumes, and the
latter are not changed in any way themselves and are not endowed with any
new additional structure.

In part V I discuss the five-vector generalizations of affine connection
and gauge fields. I then give definition to the exterior derivative of
nonscalar-valued five-vector forms and consider the five-vector analogs
of the field strength tensor. In conclusion of that part I briefly comment
on the nonspacetime analogs of five-vectors.

In part VI I first define the bivector derivative for four-vector and
four-tensor fields in the case of arbitrary Riemannian geometry. I then
define this derivative for five-vector and five-tensor fields, examine the
bivector analogs of the Riemann tensor, and introduce the notion of a
commutator for the fields of five-vector bivectors. After that I examine a
more general case of five-vector affine connection, introduce the five-vector
analog of the curvature tensor, discuss the canonical stress-energy and
angular momentum tensors corresponding to the five-vector generalization
of the covariant derivative, and then consider a possible five-vector
generalization of the Einstein and Kibble--Sciama equations. In conclusion,
I introduce the notion of the bivector derivative for the fields whose values
are vectors or tensors not directly related to space-time, and then consider
the corresponding gauge fields and discuss some of their properties.

Most of the material presented in parts II, IV, V, and VI can be easily
adapted to the case of arbitrary differentiable manifolds with metric. To
simplify the presentation, I do not indicate explicitly the smoothness
conditions for scalar functions and tensor fields under which the statements
formulated are valid. If necessary, these conditions can be retrieved without
any difficulty.

\vspace{2ex} \begin{flushleft}
\bf 2. Invariant formulation of the five-vector \\
\hspace{2.5ex} hypothesis and notations
\end{flushleft}
For any vector space $V$ (here I will be concerned with real vectors only)
one can consider a space of bivectors. A bivector is a wedge product of two
vectors:
\begin{displaymath} \bf
u \wedge v \equiv u \otimes v - v \otimes u,
\end{displaymath}
or a sum of such products. In the former case the bivector is called {\em
simple}. All bivectors are simple for a three-dimensional $V$. For $V$ with
dimension higher than 3, the sum of two simple bivectors may not be a simple
bivector. For example, as one knows from classical electrodynamics, a general
antisymmetric four-tensor of rank 2 is not always a wedge product of two
four-vectors. There, however, exist such {\em subsets} of simple bivectors
which are closed with respect to addition. Each such subset will be referred
to as a {\em vector space of simple bivectors}. The structure of such spaces
is described by the following theorem:
\begin{description}
  \item[Theorem:] If $\cal A$ is a vector space of simple bivectors
  constructed from elements of a vector space $V$ and ${\rm dim} \, {\cal A}
  > 3$, then there exists a nonzero vector ${\bf w} \in V$ such that each
  element of $\cal A$ can be presented in the form
\begin{displaymath}
\bf u \wedge w,
\end{displaymath}
  where $\bf u$ is some vector from $V$. For a given $\cal A$ the vector
  $\bf w$ is unique up to a normalization factor.
\end{description}
As one can see, the three-dimensional space is an exception: for it the sum
of any two bivectors is a simple bivector, but all its bivectors cannot be
presented in the form indicated in the Theorem.

Let us now consider vector spaces of simple bivectors with maximum dimension.
Such spaces will be called {\em maximal}. From the Theorem and the fact that
for any vector space $V$, the set of bivectors $\bf u \wedge w$, where
$\bf w$ is fixed and $\bf u$ runs through $V$, is a vector space of simple
bivectors, it follows that:
\begin{enumerate}
  \item For an $n$-dimensional vector space $V$ with $n \geq 4$, the
  dimension of any maximal vector space of simple bivectors is $n-1$.
  \item At $n \geq 5$, for each such maximal vector space $\cal A$ there
  exists a vector ${\bf w} \in V$ such that each element of $\cal A$ can be
  presented as $\bf u \wedge w$, where $\bf u$ is some vector from $V$, and
  any bivector of such form belongs to $\cal A$. I will call $\bf w$ a
  {\em directional} vector of $\cal A$.
  \item For a given maximal vector space of simple bivectors, the directional
  vector is unique up to an arbitrary normalization factor.
\end{enumerate}

\vspace{1ex}

We can now reformulate the second part of our assumption about five-vectors
as follows: {\em there exists a certain isomorphism between the space of
four-dimensional tangent vectors and one of the maximal vector spaces
of simple bivectors over} $V_{5}$. It should be emphasized that the meaning
of the latter statement is not that the two mentioned vector spaces are
isomorphic, which is merely a consequence of the definition of $V_{5}$,
but that it is supposed that there is given {\em one specific} isomorphism,
by means of which five-vectors are related to space-time and the origin of
which will become clear when we turn to the more sophisticated theory of
five-vectors presented in part II.

The mentioned isomorphism enables one to make a certain simplification in
terminology within the formal theory of five-vectors, which proves to be
quite convenient and which I will use in this part only. Namely, basing
on this isomorphism one can {\em identify} four-dimensional tangent vectors
with elements of the mentioned maximal vector space of simple bivectors,
which in this case will naturally be denoted as $V_{4}$, too. Thus, instead
of saying that four-vector $\bf U$ corresponds to bivector $\bf u \wedge w$,
one can simply write $\bf U = u \wedge w$.

As usual, the inner product of four-vectors will be denoted as $g$. The
nondegenerate inner product on the space of five-vectors will be denoted
with the symbol $h$. Under the above identification, the relation between
$g$ and $h$ is given by the following equation:
\begin{equation} \left. \begin{array}{l}
g({\bf u \wedge w, v \wedge w}) \\
\hspace*{8ex} = h({\bf u,v}) h({\bf w,w}) - h({\bf u,w}) h({\bf v,w}).
\end{array} \right. \end{equation}
The overall sign of $h$ is a matter of convention and for purely practical
reasons it is convenient to choose it so that $h$ would have the signature
$(+---+)$, to make its relation to $g$ simpler.

Let us now determine what kind of a directional vector $\bf w$ corresponds to
$V_{4}$. If $\bf w$ had a negative norm squared, one could always choose its
arbitrary normalization factor so that $h({\bf w,w}) = - 1$, and then select
an orthonormal basis of five-vectors with ${\bf e}_{5} = {\bf w}$. In that
case, for the four-vector basis ${\bf E}_{\mu} = {\bf e}_{\mu} \wedge
{\bf e}_{5}$ the inner product matrix would be
\begin{displaymath}
g_{\mu \nu} \equiv g({\bf E_{\mu}, E_{\nu}}) = {\rm diag}(-1,-1,+1,+1),
\end{displaymath}
and not of Lorentz type. Thus, the norm squared of $\bf w$ cannot be negative.

In a similar manner one can check that if $h({\bf w,w}) = 0$, the inner
product induced on the corresponding maximal vector space of simple bivectors
would be degenerate, so $\bf w$ cannot be a null vector either. Thus, one is
left with the only possibility that the directional vector of $V_{4}$ has a
positive norm squared.

Let us now fix our notations:
\begin{itemize}
  \item Five-vectors will be denoted with lower-case boldface Roman letters:
  $\bf u, v, w$, etc.
  \item A typical basis in $V_{5}$ will be denoted as ${\bf e}_{A}$, where
  $A$ (as all capital latin indices) runs 0, 1, 2, 3, and 5. An arbitrary
  five-vector $\bf u$ is expressed in terms of its components in a given
  basis as ${\bf u} = u^{A} {\bf e}_{A}$. One can choose a basis in $V_{5}$
  arbitrarily, but it is more convenient to select the fifth basis vector
  coinciding with one of the directional vectors.\footnote{It is not required
  that ${\bf e}_{5}$ be normalized.} Such bases will be called {\em standard}
  and will be used in all calculations.\footnote{As one can see, the basis
  vector and vector components related to the fifth dimension are labled
  with the index 5 rather than 4. This corresponds to the index convention
  used for $\gamma$-matrices, where the notation $\gamma_{4}$ is reserved for
  the timelike $\gamma$-matrix in the Pauli metric: $\gamma_{4} = i
  \gamma_{0}$. This also better suits the words ``fifth dimension'', and
  accentuates the fact that this direction in $V_{5}$ is distinguished as
  being the one that corresponds to the directional vector of $V_{4}$.}
  \item Four-vectors will be denoted with capital boldface Roman letters:
  $\bf U, V, W$, etc. One can choose a basis in $V_{4}$ arbitrarily and
  independently of the basis in $V_{5}$. However, it is more convenient to
  select it as
\begin{equation}
{\bf E}_{\mu} = {\bf e}_{\mu} \wedge {\bf e}_{5},
\end{equation}
  where $\mu$ (as all lower-case Greek indices) runs 0, 1, 2, and 3. I
  will refer to this basis as to the one {\em associated} with the basis
  ${\bf e}_{A}$ in $V_{5}$.
\end{itemize}

\vspace{2ex} \begin{flushleft} \bf
3. Algebraic properties of five-vectors \\ \vspace*{2ex}
\rm A. \it Transformations from one standard basis \\
\hspace*{2.5ex} to another
\end{flushleft}
Let ${\bf e}_{A}$ be an arbitrary standard basis in $V_{5}$ and let
${\bf e}'_{A}$ be another basis in $V_{5}$ such that
\begin{equation}
{\bf e}'_{A} = {\bf e}_{B} L^{B}_{\, A},
\end{equation}
where $L^{B}_{\, A}$ is a real nondegenerate $5 \times 5$ matrix. The
relation between the corresponding associated four-vector bases is
\begin{equation} \left. \begin{array}{lcl}
{\bf E}'_{\mu} & = & {\bf e}'_{\mu} \wedge {\bf e}'_{5} \; = \; {\bf e}_{A}
\wedge {\bf e}_{B} \, L^{A}_{\, \mu} L^{B}_{\, 5} \\ & = & {\bf E}_{\nu} \,
(L^{\nu}_{\mu} L^{5}_{5} - L^{5}_{\mu} L^{\nu}_{5}) \\ & & \; \; \; + \;
{\displaystyle \sum_{\alpha < \beta}} {\bf e}_{\alpha} \wedge {\bf e}_{\beta}
\, (L^{\alpha}_{\mu} L^{\beta}_{5} - L^{\beta}_{\mu} L^{\alpha}_{5}).
\end{array} \right. \end{equation}
If the basis ${\bf e}'_{A}$ is also standard, one should have
\begin{equation}
{\bf E}'_{\mu} = {\bf E}_{\nu} \Lambda^{\nu}_{\, \mu}
\end{equation}
for some real nondegenerate $4 \times 4$ matrix $\Lambda^{\nu}_{\, \mu}$.
Comparing (7) and (8), one finds that
\begin{equation}
L^{\alpha}_{\, \mu} L^{\beta}_{\, 5} - L^{\beta}_{\, \mu}
L^{\alpha}_{\, 5} = 0,
\end{equation} \begin{equation}
L^{\nu}_{\, \mu} L^{5}_{\, 5} - L^{5}_{\, \mu} L^{\nu}_{\, 5}
= \Lambda^{\nu}_{\, \mu}.
\end{equation}
Equation (9) is equivalent to the requirement
\begin{equation}
L^{\alpha}_{\, 5} = 0 \, \mbox{ for all } \, \alpha,
\end{equation}
which thus is a necessary condition of ${\bf e}'_{A}$ being a standard
basis. This is also a sufficient condition, since according to it,
\begin{displaymath}
{\bf e}'_{\, 5} = {\bf e}_{A} L^{A}_{\, 5} = {\bf e}_{5} L^{5}_{\, 5},
\end{displaymath}
and $L^{5}_{\, 5}$ cannot be zero because $L^{A}_{\, B}$ is nondegenerate.

>From (10) and (11) one obtains the formula
\begin{equation}
\Lambda^{\nu}_{\, \mu} = L^{5}_{\, 5} L^{\nu}_{\, \mu},
\end{equation}
which relates $\Lambda^{\nu}_{\, \mu}$ to $L^{\nu}_{\, \mu}$. One should
also note that
\begin{displaymath}
L^{5}_{\, 5} (L^{-1})^{5}_{\, 5} = 1 \; \; \; {\rm and} \; \; \;
(L^{-1})^{\alpha}_{\, 5} = 0,
\end{displaymath}
where $(L^{-1})^{A}_{\, B}$ is the inverse of $L^{A}_{\, B}$.

It is convenient to distinguish three different types of transformations
from one standard basis in $V_{5}$ to another:

($i$) transformations of the form
\begin{displaymath} \left \{ \begin{array}{l}
L^{5}_{\, 5} = a, \; L^{\alpha}_{\, 5} = 0 \\
L^{5}_{\, \beta} = 0, \; L^{\alpha}_{\, \beta} = a^{-1} \,
\delta^{\alpha}_{\beta} \; \; (a \neq 0),
\end{array} \right. \end{displaymath}
which will be referred to as $U$-{\em transformations};

($ii$) transformations of the form
\begin{displaymath} \left \{ \begin{array}{l}
L^{5}_{\, 5} = 1, \; L^{\alpha}_{\, 5} = 0 \\
L^{5}_{\, \beta} = a_{\beta}, \;
L^{\alpha}_{\, \beta} = \delta^{\alpha}_{\beta},
\end{array} \right. \end{displaymath}
which will be referred to as $P$-{\em transformations}; and

($iii$) transformations of the form
\begin{displaymath} \left \{ \begin{array}{l}
L^{5}_{\, 5} = 1, \; L^{\alpha}_{\, 5} = 0 \\
L^{5}_{\, \beta} = 0, \; L^{\alpha}_{\, \beta} = t^{\alpha}_{\beta},
\end{array} \right. \end{displaymath}
which will be referred to as $M$-{\em transformations} ($t^{\alpha}_
{\, \beta}$ is some nondegenerate $4 \times 4$ matrix). An arbitrary
transformation from one standard basis to another can be presented as a
composition of a $U$-, a $P$-, and an $M$-transformation. It is a simple
matter to see that $U$- and $P$-transformations have no effect on
four-vectors, i.e.\ that they induce identity transformations in $V_{4}$.
For $M$-transformations one evidently has $\Lambda^{\alpha}_{\, \beta} =
L^{\alpha}_{\, \beta} = t^{\alpha}_{\, \beta}$.

\vspace{2ex} \begin{flushleft}
\rm B. \it Symmetries and other special transformations
\end{flushleft}
If one considers $V_{5}$ by itself and takes into account the
five-orientation by introducing a Levi-Civita type tensor $\epsilon_{ABCDE}$,
the group of isomorphisms of $V_{5}$ will be SO(3,2). This symmetry is broken
when one of the maximal vector spaces of simple bivectors over $V_{5}$ is
identified with the space of four-vectors. The symmetry group of the
structure as a whole ($V_{5}$ plus $V_{4}$) is apparently SO(3,1), and the
corresponding isomorphisms are $M$-transformations (which in this case should
be interpreted in the active sense) with $t^{\alpha}_{\, \beta} \in$ SO(3,1).

One may notice that the latter transformations and $P$-transformations make
up a group isomorphic to the Poincare group. This can be easily seen by
comparing the formulae for $P$- and $M$-transformations at $t^{\alpha}_{\,
\beta} \in$ SO(3,1) with the formulae for the Poincare transformation of
covariant Lorentz coordinates in the five-dimensional representation (see
Appendix) and observing that they are identical in form. This coincidence
is not accidental. It turns out that the rules of parallel transport for
five-vectors are such that with any Lorentz coordinate system in flat
space-time one can associate either an orthonormal set of basis five-vector
fields (everywhere $h({\bf e}_{A},{\bf e}_{B}) = \eta_{AB}$), which, however,
cannot be chosen self-parallel, or a set of self-parallel basis fields
(everywhere $\nabla {\bf e}_{A} = {\bf 0}$), which can be made orthonormal
only at one point in space-time, for example, at the origin of the coordinate
system. As one will see in section 4, the elements of the self-parallel basis
at a given point transform nontrivially under space-time translations,
and in the general case the Poincare transformation of such a basis is a
composition of a certain $M$-transformation with $t^{\alpha}_{\, \beta}
\in$ SO(3,1) and a certain $P$-transformation.

\vspace{2ex} \begin{flushleft}
\rm C. \it Relation between four- and five-vector bases
\end{flushleft}
For any five-vector basis ${\bf e}_{A}$ one can construct the corresponding
associated basis of four-vectors: ${\bf E}_{\alpha} = {\bf e}_{\alpha} \wedge
{\bf e}_{5}$. It is evident that this correspondence is not mutually unique:
for any basis of five-vectors obtained from ${\bf e}_{A}$ by arbitrary $U$-
and $P$-transformations the associated basis of four-vectors will be exactly
the same. One can distinguish between all these five-vector bases only by
imposing additional requirements. One particular way of choosing the
five-vector basis for a given basis of four-vectors is based on the
following two lemmas:
\begin{description}
  \item[Lemma 1:] For any orthonormal basis of four-vectors
  ${\bf E}_{\alpha}$, there exists an orthonormal standard basis of
  five-vectors ${\bf e}_{A}$ such that ${\bf e}_{\alpha} \wedge {\bf e}_{5}
  = {\bf E}_{\alpha}$. This five-vector basis is unique up to a common sign
  of all ${\bf e}_{A}$.
\end{description}
{\em Proof} : Since all ${\bf E}_{\alpha}$ are elements of one maximal vector
space of simple bivectors, they can be presented as ${\bf E}_{\alpha} =
{\bf e}'_{\alpha} \wedge {\bf e}'_{5}$, where ${\bf e}'_{5}$ is a directional
vector of this maximal vector space and ${\bf e}'_{\alpha}$ are certain
five-vectors. One can easily show that the five vectors ${\bf e}'_{A}$ are
linearly independent and therefore form a standard basis in $V_{5}$. Let us
construct a new basis according to the formulae
\begin{displaymath} \left. \begin{array}{l}
{\bf e}_{\alpha} = (h_{5'5'})^{1/2} \{ {\bf e}'_{\alpha} -
(h_{\alpha'5'})/(h_{5'5'}) \, {\bf e}'_{5} \} \\
{\bf e}_{5} = (h_{5'5'})^{-1/2} {\bf e}'_{5},
\end{array} \right. \end{displaymath}
where $h_{A'B'} \equiv h({\bf e}'_{A},{\bf e}'_{B})$. This is also a standard
basis, and simple calculations show that ${\bf e}_{\alpha} \wedge {\bf e}_{5}
= {\bf E}_{\alpha}$ and $h({\bf e}_{A},{\bf e}_{B}) = \eta_{AB}$, so it has
been demonstrated that the required basis exists.

If ${\bf e}''_{A}$ is another basis that satisfies the same requirements as
${\bf e}_{A}$, and ${\bf e}''_{A} = {\bf e}_{B} L^{B}_{\, A}$, then one can
easily show that $L^{5}_{\, 5} = \pm 1$, $L^{\alpha}_{\, 5} =
L^{5}_{\, \beta} = 0$, and $L^{\alpha}_{\, \beta} = (L^{5}_{\, 5})^{-1}
\delta^{\alpha}_{\beta}$, so either ${\bf e}''_{A} = {\bf e}_{A}$ or
${\bf e}''_{A} = - {\bf e}_{A}. \; \;$\rule{0.8ex}{1.7ex}

\vspace{3ex}

We thus see that for the special case of an {\em orthonormal} four-vector
basis one can fix the corresponding five-vector basis up to a sign by
requiring that the latter be orthonormal, too. In a certain sense, this is
a natural choice. It is also natural that the orthonormality condition does
not fix the overall sign of the basis five-vectors, since this sign has no
effect on their inner products.\footnote{To fix the five-vector basis
unambiguously, one has to impose one more requirement. For example, one may
observe that by changing the overall sign of the basis five-vectors one
changes the five-orientation of the basis, so one can fix a single basis by
requiring that $\epsilon_{01235} = +1$ or that $\epsilon_{01235} = -1$.}
In the general case, the selection of the five-vector basis can be based on
the following lemma:
\begin{description}
   \item[Lemma 2:] For an arbitrary basis of four-vectors ${\bf E}_{\alpha}$,
   there exists a standard basis of five-vectors ${\bf e}_{A}$ such that
   $h({\bf e}_{5},{\bf e}_{5}) = 1$, $h({\bf e}_{5},{\bf e}_{\alpha}) = 0$,
   and ${\bf e}_{\alpha} \wedge {\bf e}_{5} = {\bf E}_{\alpha}$. This
   five-vector basis is unique up to a common sign of all ${\bf e}_{A}$.
\end{description}
{\em Proof} : It is evident that there exists a matrix
$\Lambda^{\alpha}_{\, \beta}$ such that ${\bf E}'_{\alpha} = {\bf E}_{\beta}
\Lambda^{\beta}_{\, \alpha}$ is an orthonormal basis in $V_{4}$. According
to Lemma 1, there exists an orthonormal five-vector basis ${\bf e}'_{A}$
such that ${\bf e}'_{\alpha} \wedge {\bf e}'_{5} = {\bf E}'_{\alpha}$. One
can easily check that the basis
\begin{displaymath}
{\bf e}_{\alpha} = {\bf e}'_{\beta} (\Lambda^{-1})^{\beta}_{\, \alpha} \;
\mbox{ and } \; {\bf e}_{5} = {\bf e}'_{5}
\end{displaymath}
is such that $h({\bf e}_{5},{\bf e}_{5}) = 1$, $h({\bf e}_{5},
{\bf e}_{\alpha}) = 0$, and ${\bf e}_{\alpha} \wedge {\bf e}_{5} =
{\bf E}_{\alpha}$, so it has been demonstrated that the required basis
exists.

If ${\bf e}''_{A}$ is another basis that satisfies the same requirements as
${\bf e}_{A}$, one can construct the basis ${\bf e}'''_{5} = {\bf e}''_{5}$,
${\bf e}'''_{\alpha} = {\bf e}''_{\beta} \Lambda^{\beta}_{\, \alpha}$ and
check that ${\bf e}'''_{A}$ is orthonormal and that ${\bf e}'''_{\alpha}
\wedge {\bf e}'''_{5} = {\bf E}'_{\alpha}$. Thus, by virtue of Lemma 1,
one has ${\bf e}'''_{A} = \pm {\bf e}'_{A}$, so
\begin{displaymath} \hspace*{3ex} \left. \begin{array}{l}
{\bf e}''_{5} = {\bf e}'''_{5} = \pm {\bf e}'_{5} = \pm {\bf e}_{5} \\
{\bf e}''_{\alpha} = {\bf e}'''_{\beta} (\Lambda^{-1})^{\beta}_{\, \alpha}
= \pm {\bf e}'_{\beta} (\Lambda^{-1})^{\beta}_{\, \alpha} =
\pm {\bf e}_{\alpha}. \hspace*{5ex} \rule{0.8ex}{1.7ex}
\end{array} \right. \end{displaymath}

\vspace{3ex}

A standard five-vector basis that satisfies the requirements $h({\bf e}_{5},
{\bf e}_{5}) = 1$ and $h({\bf e}_{5},{\bf e}_{\alpha}) = 0$ will be called a
{\em regular} basis. Thus, Lemma 2 states that for a given four-vector basis
there exist but two corresponding regular five-vector bases, differing from
each other only in the overall sign of the basis five-vectors. A regular
basis is very convenient since in it
\begin{displaymath}
h_{5 5} = 1, \; h_{\alpha 5} = 0, \mbox{ and } h_{\alpha \beta} =
g_{\alpha \beta},
\end{displaymath}
which simplifies algebraic transformations, and (if one chooses the
five-vector basis this way at every point)
\begin{displaymath}
\partial_{\mu} h_{5 5} = \partial_{\mu} h_{\alpha 5} = 0 \; \mbox{ and } \;
\partial_{\mu} h_{\alpha \beta} = \partial_{\mu} g_{\alpha \beta},
\end{displaymath}
which is convenient when one evaluates the derivatives.

\vspace{2ex} \begin{flushleft}
\bf 4. Differential properties of five-vectors \\
\vspace*{2ex}
\rm A. \it Relation between parallel transports \\
\hspace{2.5ex} of four- and five-vectors
\end{flushleft}
When considering the differential properties of five-vectors, one should
imagine that at each point in space-time there exists a tangent space of
five-vectors. As for any other type of vector-like objects considered in
space-time, one can speak of parallel transport of five-vectors from one
point to another. It seems natural to suppose that the rules of this
transport should be related in some way to similar rules for four-dimensional
tangent vectors. It is obvious that this relation cannot be derived from
algebraic properties of five-vectors, and to obtain it one has to make some
additional assumption about five-vectors, which ought to be regarded as part
of their definition.

The simplest and the most natural form of the relation in question can be
obtained by postulating that parallel transport preserves the isomorphism
between the space of four-vectors and one of the maximal vector spaces of
simple bivectors over $V_{5}$, which has been discussed above. A more
precise formulation of this statement is the following:
\begin{equation} \begin{minipage}{40ex} \it
If four-vector $\bf U$ corresponds to bivector $\bf u \wedge w$, then
the transported $\bf U$ corresponds to the transported $\bf u \wedge w$.
\end{minipage} \end{equation}
This assumption has two consequences, which can be conveniently expressed in
terms of connection coefficients. Let us define the latter for five-vectors
as
\begin{displaymath}
\nabla_{\mu} {\bf e}_{A} = {\bf e}_{B} G^{B}_{\, A \mu},
\end{displaymath}
where $\nabla_{\mu} \equiv \nabla_{{\bf E}_{\mu}}$ denotes the covariant
derivative in the direction of the basis four-vector ${\bf E}_{\mu}$. The
connection coefficients for four-vectors will be denoted in the standard way:
\begin{equation}
\nabla_{\mu} {\bf E}_{\alpha}={\bf E}_{\beta} \Gamma^{\beta}_{\, \alpha \mu}.
\end{equation}
In the usual manner one can obtain the expression for the components of the
covariant derivative of an arbitrary five-vector field ${\bf u}$:
\begin{displaymath}
\nabla_{\mu} {\bf u} = \nabla_{\mu} ( u^{A} {\bf e}_{A} )
= (\partial_{\mu} u^{A} + G^{A}_{\, B \mu} u^{B}) \, {\bf e}_{A}
\equiv u^{A}_{\; \; ; \, \mu} {\bf e}_{A},
\end{displaymath}
and the transformation formula for five-vector connection coefficients
corresponding to the transformations ${\bf E}'_{\mu} = {\bf E}_{\nu}
\Lambda^{\nu}_{\, \mu}$ and ${\bf e}'_{A} = {\bf e}_{B} L^{B}_{\, A}$ of
the four- and five-vector bases:
\begin{displaymath}
G'^{A}_{\; B \mu} =
(L^{-1})^{A}_{\; C} G^{C}_{\; D \nu} L^{D}_{\; B} \Lambda^{\nu}_{\; \mu}
+ (L^{-1})^{A}_{\; C} ( \partial_{\nu} L^{C}_{\; B} ) \Lambda^{\nu}_{\; \mu}.
\end{displaymath}

If at each point the five-vector basis ${\bf e}_{A}$ is chosen standard and
${\bf E}_{\alpha}$ is the associated basis of four-vectors, then
\begin{equation} \left. \begin{array}{rcccl}
\nabla_{\mu} {\bf E}_{\alpha} \! & = \! & \nabla_{\mu} ( {\bf e}_{\alpha}
\wedge {\bf e}_{5} ) & & \\ & = \! & (\nabla_{\mu} {\bf e}_{\alpha}) \wedge
{\bf e}_{5} \hspace{-1ex} & + \hspace{-1ex} & {\bf e}_{\alpha} \wedge
(\nabla_{\mu} {\bf e}_{5}) \\ & = \! & {\bf e}_{\beta} \wedge {\bf e}_{5}
\, (G^{\beta}_{\, \alpha \mu} \hspace{-1ex} & + \hspace{-1ex} &
\delta^{\beta}_{\alpha} G^{5}_{\, 5 \mu}) \\ & & & + \hspace{-1ex} &
{\bf e}_{\alpha} \wedge {\bf e}_{\beta} \, G^{\beta}_{\, 5 \mu}.
\end{array} \right. \end{equation}
Comparing (14) and (15) one finds that
\begin{equation}
G^{\alpha}_{\, 5 \mu} = 0 \mbox{ for all } \alpha,
\end{equation}
and
\begin{equation}
\Gamma^{\alpha}_{\, \beta \mu} = G^{\alpha}_{\, \beta \mu}
+ \delta^{\alpha}_{\beta} G^{5}_{\, 5 \mu}.
\end{equation}
These equations express the relation between the rules of parallel transport
for four- and five-vectors. One should notice that they tell one nothing
about the coefficients $G^{5}_{\, \beta \mu}$, so as far as four-vector
parallel transport is concerned, the latter can be absolutely arbitrary.

\vspace{2ex} \begin{flushleft}
\rm B. \it Five-vectors in flat space-time
\end{flushleft}
It is a well known fact that owing to its special geometric features, flat
space-time possesses a symmetry which in application to scalars, four-vectors
and other four-tensors can be formulated as the following principle:
\begin{quote}
For any set of scalar, four-vector and four-tensor fields in flat space-time,
by means of a certain procedure one can construct a new set of fields (which
will be called {\em equivalent}) such that at each point in space-time these
new fields satisfy the same algebraic and differential relations that the
original fields satisfy at a certain corresponding point.
\end{quote}
The procedure by means of which the equivalent fields are constructed can be
formulated as follows:
\begin{enumerate}
   \item Introduce a system of Lorentz coordinates $x^{\alpha}$. \\
   Introduce the corresponding coordinate four-vector basis ${\bf E}_{\alpha}
    = \partial / \partial x^{\alpha}$. \\
   Introduce the corresponding bases for all other four-tensors.
   \item Each scalar field $f$ will then determine and be determined by one
   real coordinate function $f(x)$. \\
   Each four-vector field $\bf U$ will determine and be determined by four
   real coordinate functions $U^{\alpha}(x)$ (= components of $\bf U$ in
   the basis ${\bf E}_{\alpha}$). \\
   Each four-tensor field $\bf T$ will determine and be determined by an
   appropriate number of real coordinate functions $T^{\alpha \beta \ldots
   \gamma}_{\lambda \nu \ldots \mu}(x)$ (= components of $\bf T$ in the
   appropriate tensor basis corresponding to ${\bf E}_{\alpha}$).
   \item Introduce a new system of Lorentz coordinates $x'^{\alpha}$. \\
   Introduce the corresponding four-vector basis ${\bf E}'_{\alpha} =
   \partial / \partial x'^{\alpha}$. \\
   Introduce the corresponding  bases for all other four-tensors.
   \item Then the equivalent scalar, four-vector and four-tensor fields will
   be determined in the new coordinates and new bases by the {\em same
   functions} $f(\cdot)$, $U^{\alpha}(\cdot)$, \ldots , $T^{\alpha \beta
   \ldots \gamma}_{\lambda \nu \ldots \mu}(\cdot)$ that determine the
   original fields in the old coordinates and old bases.
\end{enumerate}

The above symmetry principle and the corresponding procedure for constructing
equivalent fields follow from the definition of flat space-time and the
assumptions that in it $\nabla$ is torsion-free and satisfies the condition
of compatibility with metric:
\begin{equation}
\nabla g = 0.
\end{equation}
The latter two assumptions enable one to find the rules of parallel transport
for four-vectors, and knowing these one can {\em prove} that the above
symmetry principle holds. For five-vectors let us reverse the problem: let
us {\em suppose} that only those five-vectors have anything to do with
reality for which there holds a symmetry principle similar to the one
formulated above for four-vectors, and then use this principle to determine
the rules of parallel transport for five-vectors in flat space-time.

Let us introduce a system of Lorentz coordinates, $x^{\alpha}$, and consider
the following set of fields:
\begin{enumerate}
 \item Four four-vector basis fields ${\bf E}_{\alpha} = \partial / \partial
       x^{\alpha}$.
 \item Five {\em continuous} five-vector fields ${\bf e}_{A}$ such that at
       each point they make up a {\em regular} basis corresponding to
       ${\bf E}_{\alpha}$.\footnote{There are {\em two} sets of fields
       like that (see Lemma 2) and we choose one of them.}
 \item $5 \times 5 \times 4 = 100$ scalar fields $H^{A}_{\, B \mu}$ such that
       everywhere
\begin{displaymath}
\nabla_{\mu} {\bf e}_{A} = {\bf e}_{B} H^{B}_{\, A \mu}.
\end{displaymath} \end{enumerate}
By their definition, $H^{A}_{\, B \mu}$ are connection coefficients for the
basis fields ${\bf e}_{A}$, and since all these bases are standard, one
should have $H^{\alpha}_{\, 5 \mu} = 0$. Furthermore, since space-time is
flat and ${\bf E}_{\alpha}$ is a Lorentz basis, the corresponding four-vector
connection coefficients are zero, so one should have
\begin{equation}
H^{\alpha}_{\, \beta \mu} + \delta^{\alpha}_{\beta} H^{5}_{\, 5 \mu} =
\Gamma^{\alpha}_{\, \beta \mu} = 0.
\end{equation}

Let us now consider another system of Lorentz coordinates,
$x'^{\alpha}$, such that
\begin{equation}
x'^{\alpha} = x^{\alpha} + a^{\alpha},
\end{equation}
where $a^{\alpha}$ are four arbitrary constant parameters. The fields
equivalent to ${\bf E}_{\alpha}$ are ${\bf E}_{\alpha}$ themselves, since by
virtue of the symmetry principle,
\begin{displaymath}
{\bf E}'_{\alpha} = \partial / \partial x'^{\alpha}
 = \partial / \partial x^{\alpha} = {\bf E}_{\alpha}.
\end{displaymath}
In view of this, for the fields equivalent to ${\bf e}_{A}$ one has only
two options: either ${\bf e}'_{A} = {\bf e}_{A}$ or ${\bf e}'_{A} = - \,
{\bf e}_{A}$. Since coordinate transformation (20) depends continuously on
$a^{\alpha}$, it is natural to require that the same be true of the
corresponding field transformation, which leaves us with only one
possibility: ${\bf e}'_{A} = {\bf e}_{A}$.

Finally, by virtue of the symmetry principle, the scalar fields equivalent to
$H^{A}_{\, B \mu}$ are such that
\begin{equation}
H'^{A}_{\, B \mu}(x'^{\alpha} = y^{\alpha}) = H^{A}_{\, B \mu} (x^{\alpha} =
y^{\alpha}),
\end{equation}
at all $y^{\alpha}$. Since equivalent fields must satisfy the same relations
as the original fields, one should have
\begin{displaymath}
{\bf e}_{B} H'^{B}_{\; A \mu} = {\bf e}'_{B} H'^{B}_{\; A \mu}
 = \nabla_{{\bf E}'_{\mu}} {\bf e}'_{A} = \nabla_{{\bf E}_{\mu}} {\bf e}_{A}
 = {\bf e}_{B} H^{B}_{\, A \mu}.
\end{displaymath}
Thus, at any point $Q$ one has
\begin{displaymath}
H'^{A}_{\; B \mu}(Q) = H^{A}_{\, B \mu}(Q).
\end{displaymath}
Comparing this with equation (21), one finds that at any $a^{\alpha}$
\begin{displaymath} \left. \begin{array}{rcl}
H^{A}_{\, B \mu}(x^{\alpha} = 0) & = & H'^{A}_{\; B \mu} (x^{\alpha} = 0) \\
& = & H'^{A}_{\; B \mu}(x'^{\alpha} = a^{\alpha}) \\ & = &
H^{A}_{\, B \mu}(x^{\alpha} = a^{\alpha}),
\end{array} \right. \end{displaymath}
which means that each $H^{A}_{\, B \mu}$ is a constant scalar field.

Let us consider a third system of Lorentz coordinates, $x''^{\alpha}$, which
are related to $x^{\alpha}$ as
\begin{displaymath}
x''^{\alpha} = \Lambda^{\alpha}_{\; \beta} x^{\beta},
\end{displaymath}
where $\Lambda^{\alpha}_{\; \beta}$ is an arbitrary constant matrix from
SO(3,1). By virtue of the symmetry principle, the fields equivalent to
${\bf E}_{\alpha}$ are
\begin{displaymath}
{\bf E}''_{\alpha} = \partial / \partial x''^{\alpha}
 = \partial / \partial x^{\beta} \, (\Lambda^{-1})^{\beta}_{\; \alpha}
 = {\bf E}_{\beta} \, (\Lambda^{-1})^{\beta}_{\; \alpha} .
\end{displaymath}
It is a simple matter to check that if one requires the field transformation
to depend continuously on parameters $\Lambda^{\alpha}_{\; \beta}$, the
fields equivalent to ${\bf e}_{A}$ will be
\begin{displaymath}
{\bf e}''_{\alpha} = {\bf e}_{\beta} \, (\Lambda^{-1})^{\beta}_{\; \alpha}
\; \mbox{ and } \; {\bf e}''_{5} = {\bf e}_{5}.
\end{displaymath}
Finally, since it has been found that each $H^{A}_{\; B \mu}$ is a constant
scalar field, one should have
\begin{displaymath}
H''^{A}_{\, \; B \mu} = H^{A}_{\, B \mu}.
\end{displaymath}
Since equivalent fields satisfy the same relations,
\begin{displaymath} \begin{array}{rl}
{\bf e}_{5} H^{5}_{\; 5 \mu} \! & = {\bf e}''_{5} H^{5}_{\; 5 \mu}
= {\bf e}''_{B} H^{B}_{\; 5 \mu} = {\bf e}''_{B} H''^{B}_{\, \; 5 \mu} \\
& = \nabla_{{\bf E}''_{\mu}} {\bf e}''_{5} = \nabla_{{\bf E}_{\nu}}
{\bf e}_{5} \, (\Lambda^{-1})^{\nu}_{\; \mu} \\
&  = {\bf e}_{B} H^{B}_{\, 5 \nu} \, (\Lambda^{-1})^{\nu}_{\; \mu}
= {\bf e}_{5} H^{5}_{\, 5 \nu} \, (\Lambda^{-1})^{\nu}_{\; \mu},
\end{array} \end{displaymath}
so one should have
\begin{displaymath}
H^{5}_{\; 5 \mu} = H^{5}_{\, 5 \nu} \, (\Lambda^{-1})^{\nu}_{\; \mu}
\end{displaymath}
for all $\Lambda^{\mu}_{\; \nu}$ from SO(3,1), which is only possible
if $H^{5}_{\; 5 \mu} = 0$. From equation (19) it then follows that
$H^{\alpha}_{\; \beta \mu} = 0$ for all $\alpha$, $\beta$, and $\mu$.
Finally, one has
\begin{displaymath} \begin{array}{rl}
{\bf e}_{5} H^{5}_{\; \alpha \mu} \! & = {\bf e}''_{5} H^{5}_{\; \alpha \mu}
= {\bf e}''_{B} H^{B}_{\; \alpha \mu} = {\bf e}''_{B} H''^{B}_{\; \;
\alpha \mu} \\ & = \nabla_{{\bf E}''_{\mu}} {\bf e}''_{\alpha} =
\nabla_{{\bf E}_{\nu}} {\bf e}_{\beta} \, (\Lambda^{-1})^{\beta}_{\; \alpha}
(\Lambda^{-1})^{\nu}_{\; \mu} \\ & = {\bf e}_{A} H^{A}_{\, \beta \nu} \,
(\Lambda^{-1})^{\beta}_{\; \alpha} \, (\Lambda^{-1})^{\nu}_{\; \mu} \\ &
= {\bf e}_{5} H^{5}_{\, \beta \nu} \, (\Lambda^{-1})^{\beta}_{\; \alpha} \,
(\Lambda^{-1})^{\nu}_{\; \mu},
\end{array} \end{displaymath}
so one should have
\begin{displaymath}
H^{5}_{\, \alpha \mu} = H^{5}_{\, \beta \nu}
\, (\Lambda^{-1})^{\beta}_{\; \alpha} \, (\Lambda^{-1})^{\nu}_{\; \mu}
\end{displaymath}
for all $\Lambda^{\mu}_{\; \nu}$ from SO(3,1). This is only possible if
$H^{5}_{\; \alpha \mu}$ is proportional to the Minkowski metric tensor,
$\eta_{\alpha \mu}$. Denoting the proportionality factor (which should be a
constant since $H^{5}_{\; \alpha \mu}$ are constant fields) as $-\kappa$,
one can summarize our findings about $H^{A}_{\; B \mu}$ as follows:
\begin{equation}
H^{\alpha}_{\; \beta \mu} =  H^{\alpha}_{\; 5 \mu} = H^{5}_{\; 5 \mu} = 0
  \; \mbox{ and } \; H^{5}_{\; \beta \mu} = - \, \kappa \eta_{\beta \mu}.
\end{equation}
Thus, any orthonormal set of continuous five-vector basis fields
${\bf e}_{A}$ associated with a Lorentz four-vector basis in flat space-time
satisfy the following differential equations:
\begin{equation}
\nabla_{\mu} {\bf e}_{5} = 0 \; \mbox{ and } \; \nabla_{\mu} {\bf e}_{\alpha}
 = - \, \kappa \eta_{\alpha \mu} \, {\bf e}_{5},
\end{equation}
where $\kappa$ is a constant, which cannot be found from symmetry
considerations. These equations determine the rules of parallel
transport for five-vectors in flat space-time.

\vspace{2ex} \begin{flushleft}
\rm C. \it Equation for $h$
\end{flushleft}
Let us now express the contents of equation (23) in an equivalent form: as
an equation for the first covariant derivative of the inner product $h$
regarded as a five-tensor. From equations (22) and the fact that in the
orthonormal basis ${\bf e}_{A}$ introduced in the previous subsection
$h_{AB} = \eta_{AB}$ at every point, it follows that
\begin{displaymath} \begin{array}{l}
h_{55;\mu} = \partial_{\mu} h_{55} - h_{A5} H^{A}_{\; 5 \mu} - h_{5B}
H^{B}_{\; 5 \mu} = 0, \\ h_{\alpha 5;\mu} = \partial_{\mu} h_{\alpha 5}
- h_{A5} H^{A}_{\; \alpha \mu} - h_{\alpha B} H^{B}_{\; 5 \mu} \\
\hspace{24ex} = - \; h_{55} H^{5}_{\; \alpha \mu} = \kappa \eta_{\alpha \mu},
\\ h_{\alpha \beta ;\mu} = \partial_{\mu} h_{\alpha \beta} - h_{A \beta}
H^{A}_{\; \alpha \mu} - h_{\alpha B} H^{B}_{\; \beta \mu} = 0.
\end{array} \end{displaymath}
These equations can be presented in the following covariant form:
\begin{equation} \begin{array}{l}
h_{55 ; \mu} = 0, \; \; h_{\alpha 5 ; \mu} = \kappa g_{\alpha \mu} \\
h_{55} h_{\alpha \beta ; \mu} = \kappa (g_{\alpha \mu} h_{\beta 5} +
g_{\beta \mu} h_{\alpha 5} ),
\end{array} \end{equation}
which is the same in any standard five-vector basis. It is not difficult to
see that equations (24) are components of the following abstract equation:
\begin{equation} \begin{array}{l}
h({\bf e,e}) \{ \nabla_{\bf U} h \} ({\bf v,w}) \\
\hspace{8ex} = \kappa g({\bf U}, {\bf v} \wedge {\bf e}) h({\bf w,e}) \\
\hspace{16ex} + \; \kappa g({\bf U}, {\bf w} \wedge {\bf e}) h({\bf v,e}),
\end{array} \end{equation}
where $\{ \nabla_{\bf U} h \}({\bf v,w}) \equiv \partial_{\bf U} h({\bf v,w})
- h(\nabla_{\bf U}{\bf v,w}) - h({\bf v},\nabla_{\bf U}{\bf w})$ is the
covariant derivative of the tensor $h$; $\bf v$ and $\bf w$ are any two
five-vector fields; $\bf U$ is an arbitrary four-vector; and $\bf e$ is a
directional vector of $V_{4}$ (which is not required to be normalized).

Equation (25) establishes a relation between the Riemannian geometry of
space-time, represented by the inner product $h$, and the rules of parallel
transport for five-vectors. At $\kappa = 0$ it acquires a form similar to
that of equation (18) for the four-dimensional metric tensor:
\begin{displaymath}
\nabla h = 0,
\end{displaymath}
and can be given a similar simple interpretation: that the inner product of
two five-vectors is invariant under parallel transport. Equation (25) at
$\kappa \neq 0$ will be discussed further in part II.

Let us now examine more closely the properties of five-vectors in flat
space-time.

\vspace{2ex} \begin{flushleft}
\rm D. \it Self-parallel basis
\end{flushleft}
Any set of Lorentz basis four-vector fields in flat space-time
has two special features: it is orthonormal (everywhere
$g_{\alpha \beta} = \eta_{\alpha \beta}$) and self-parallel
(everywhere $\Gamma^{\alpha}_{\; \beta \mu} = 0$). This fact is closely
related to equation (18) for the metric tensor $g$: if $\nabla g$ were
nonzero, a basis like that could not exist.

With five-vectors one has a similar situation at $\kappa = 0$: as one
can see from formulae (22), the orthonormal basis ${\bf e}_{A}$ is then
self-parallel and, accordingly, the first covariant derivative of $h$ is
identically zero, as is seen from equation (25).

The situation is different at $\kappa \neq 0$. Since $\nabla h$ is nonzero,
the requirements of orthonormality and self-parallelism become conflicting
in the sense that one can have either orthonormality or self-parallelism
but not both at the same time.

The basis ${\bf e}_{A}$ of subsection B is orthonormal by definition but is
not self-parallel, as is seen from equations (22) or (23). In the following
I will call it an $O$-basis (`$O$' stands for `{\em orthonormal}')
associated with a given system of Lorentz coordinates $x^{\mu}$. Let us now
construct a self-parallel basis, ${\bf p}_{A}$, that would coincide with
${\bf e}_{A}$ at the origin of the considered coordinate system. Being a
self-parallel basis, ${\bf p}_{A}$ should satisfy the following differential
equations:
\begin{displaymath}
\nabla_{\mu} {\bf p}_{A} = 0.
\end{displaymath}
If ${\bf p}_{A} = {\bf e}_{B} N^{B}_{\, A}$, then
\begin{displaymath}
\nabla_{\mu} {\bf p}_{A} = \nabla_{\mu} ({\bf e}_{B} N^{B}_{\, A})
= {\bf e}_{B} (\partial_{\mu} N^{B}_{\, A} + H^{B}_{\, C \mu} N^{C}_{\, A}),
\end{displaymath}
where $H^{A}_{\, B \mu}$ are given by equations (22). Considering that
${\bf p}_{A}$ and ${\bf e}_{A}$ should coincide at $x = 0$, one obtains
the following system of equations for the 25 scalar coordinate functions
$N^{A}_{\, B}(x)$:
\begin{displaymath}
\partial_{\mu} N^{A}_{\, B}(x) + H^{A}_{\, C \mu} N^{C}_{\, B}(x) = 0 \;
\mbox { and } \; N^{A}_{\, B}(0) = \delta^{A}_{B}.
\end{displaymath}
This system can be easily solved and gives
\begin{displaymath} \begin{array}{l}
N^{5}_{\, 5}(x) = 1, \; \; N^{\alpha}_{\, \beta}(x) = \delta^{\alpha}_{\beta},
\\ N^{\alpha}_{\, 5}(x) = 0, \; \; N^{5}_{\, \alpha}(x) = \kappa x_{\alpha},
\end{array} \end{displaymath}
where $x_{\alpha} \equiv \eta_{\alpha \beta} x^{\beta}$ are covariant Lorentz
coordinates. We thus see that ${\bf p}_{A}$ are expressed in terms of
${\bf e}_{A}$ as follows:
\begin{equation} \begin{array}{l}
{\bf p}_{\alpha}(x) = {\bf e}_{\alpha}(x) + \kappa x_{\alpha} {\bf e}_{5}(x)
\\ {\bf p}_{5}(x) = {\bf e}_{5} (x).
\end{array} \end{equation}
I will call ${\bf p}_{A}$ a $P$-basis (`$P$' stands for `{\em parallel}')
associated with the given system of Lorentz coordinates. Simple calculations
show that
\begin{equation} \begin{array}{l}
h({\bf p}_{\alpha},{\bf p}_{\beta}) = \eta_{\alpha \beta}
+ \kappa^{2} x_{\alpha} x_{\beta} \\ h({\bf p}_{\alpha},{\bf p}_{5}) =
\kappa x_{\alpha}, \; \; h({\bf p}_{5},{\bf p}_{5}) = 1,
\end{array} \end{equation}
so ${\bf p}_{A}$ are orthonormal only at the origin.

Thus, with any system of Lorentz coordinates in flat space-time one can
associate two special sets of five-vector basis fields: an $O$-basis, which
is orthonormal everywhere but is not self-parallel, or a $P$-basis, which is
self-parallel but is not orthonormal anywhere except for the origin. At
$\kappa = 0$ the two bases coincide.

\vspace{2ex} \begin{flushleft}
\rm E. \it Poincare transformation of five-tensor \\
\hspace*{2.5ex} components
\end{flushleft}
Let us now derive the formulae for transformation of five-vector components
and of components of other five-tensors as one passes from one system of
Lorentz coordinates to another.

In the general case, with transformation of the five-vector basis according
to the formula
\begin{displaymath}
{\bf e}_{A} \rightarrow {\bf e}'_{A} = {\bf e}_{B} L^{B}_{\, A},
\end{displaymath}
the components of an arbitrary five-vector $\bf v$ transform as
\begin{equation}
v^{A} \rightarrow v'^{A} = (L^{-1})^{A}_{\, B} \, v^{B}.
\end{equation}
If $\tilde{\bf o}^{A}$ is the basis of five-vector 1-forms dual to
${\bf e}_{A}$, one should have
\begin{displaymath}
\tilde{\bf o}^{A} \rightarrow \tilde{\bf o}'^{A} = (L^{-1})^{A}_{\, B} \,
\tilde{\bf o}^{B},
\end{displaymath}
and, accordingly, the components of an arbitrary five-vector 1-form
$\widetilde{\bf w}$ in this dual basis transform as
\begin{equation}
w_{A} \rightarrow w'_{A} = w_{B} L^{B}_{\, A}.
\end{equation}

Consider now an arbitrary Poincare transformation of Lorentz coordinates:
\begin{equation}
x^{\mu} \rightarrow x'^{\mu} = \Lambda^{\mu}_{\, \nu} x^{\nu} + a^{\mu}.
\end{equation}
The same reasoning as in subsection B shows that the corresponding $O$-basis
transforms as
\begin{equation}
{\bf e}'_{\alpha} = {\bf e}_{\beta} \, (\Lambda^{-1})^{\beta}_{\, \alpha}
\; \mbox{ and } \; {\bf e}'_{5} = {\bf e}_{5},
\end{equation}
and from formulae (28) and (29) one obtains the following transformation
laws for components of five-vectors and forms:
\begin{equation} \left \{ \begin{array}{l}
v'^{\alpha} = \Lambda^{\alpha}_{\, \beta} \, v^{\beta} \\
v'^{5} = v^{5}
\end{array} \right. {\rm \; and \; \;} \left \{ \begin{array}{l}
w'_{\alpha} = w_{\beta} (\Lambda^{-1})^{\beta}_{\, \alpha} \\
w'_{5} = w_{5}.
\end{array} \right. \end{equation}
Thus, the first four components of any five-vector or five-vector 1-form
in the $O$-basis transform exactly as components of a four-vector or a
four-vector 1-form, while the fifth component behaves as a scalar.

Let us now see what happens in the $P$-basis. According to formulae (26) and
(31) and to formula (44) of Appendix, one has
\begin{displaymath}
{\bf p}'_{5} = {\bf e}'_{5} = {\bf e}_{5} = {\bf p}_{5}
\end{displaymath}
and
\begin{displaymath} \begin{array}{rcl}
{\bf p}'_{\alpha} & = & {\bf e}'_{\alpha} + \kappa x'_{\alpha} {\bf e}'_{5}
\\ & = & ({\bf e}_{\beta} + \kappa x_{\beta} {\bf e}_{5}) \, (\Lambda^{-1})
^{\beta}_{\, \alpha} + \kappa a_{\alpha} {\bf e}_{5} \\ & = & {\bf p}_{\beta}
\, (\Lambda^{-1})^{\beta}_{\, \alpha} + \kappa a_{\alpha} {\bf p}_{5},
\end{array} \end{displaymath}
where $a_{\alpha} = \eta_{\alpha \beta} a^{\beta}$. Formulae (28) and (29)
now give
\begin{flushright}
\hspace{4ex} \hfill $\left\{ \begin{array}{l} v'^{\alpha} =
\Lambda^{\alpha}_{\, \beta} v^{\beta} \\ v'^{5} = v^{5} - \kappa a_{\alpha}
\Lambda^{\alpha}_{\, \beta} \, v^{\beta} \end{array} \right.$ \hspace{4.5ex}
\hfill (33a) \end{flushright} and \begin{flushright}
\hspace{4ex} \hfill $\left\{ \begin{array}{l} w'_{\alpha} = w_{\beta}
(\Lambda^{-1})^{\beta}_{\, \alpha} + \kappa a_{\alpha} w_{5} \\ w'_{5} =
w_{5}. \end{array} \right.$ \hfill (33b)
\end{flushright} \setcounter{equation}{33}
We thus see that at $\kappa \neq 0$ the transformation laws for five-tensor
components in the $P$-basis are essentially different from what one has in
the $O$-basis. In particular, these components transform nontrivially under
space-time translations, and now one is able to understand why.

A global $P$-basis can exist only in flat space-time, where the parallel
transport of five-vectors is independent of the path along which it is
made. A $P$-basis can be constructed by choosing an orthonormal five-vector
basis at one point and transporting it parallelly to all other points in
space-time. Since (at $\kappa \neq 0$) the inner product of five-vectors
is not conserved by parallel transport, the $P$-basis cannot be orthonormal
at every point. Actually, the rules of parallel transport for five-vectors
are such that ${\bf p}_{A}$ are orthonormal only at the origin. Moreover,
as one can see from formulae (27), at each point the inner product matrix
$h_{AB} \equiv h({\bf p}_{A},{\bf p}_{B})$ has its own value, different from
the values it has at all other points. This means that having a $P$-basis,
one is able to distinguish points without using any coordinates. In fact, if
need be, one can recover the relevant Lorentz coordinates by simply
calculating the inner product of ${\bf p}_{\alpha}$ and ${\bf p}_{5}$
and using the formula
\begin{displaymath}
x_{\alpha}(Q) = \kappa^{-1} h({\bf p}_{\alpha}(Q),{\bf p}_{5}(Q)).
\end{displaymath}
Thus, the $P$-basis is a structure which is rigitly connected to space-time
points and to one of the Lorentz coordinate systems. When the latter is
changed, the $P$-basis changes too.

\vspace{2ex} \begin{flushleft}
\bf 5. Examples of five-tensors \\
\vspace*{2ex}
\rm A. \it How to find a five-vector or a five-tensor
\end{flushleft}
In the previous two sections we have examined the basic algebraic and
differential properties of five-vectors. There now arises a natural question:
are there any physical or purely geometric quantities that are described
by five-vectors or by other nontrivial five-tensors (by the ones not
reducible to a four-tensor)? This brings us to another question: how can one
discover a five-vector or a five-tensor? One possible answer to this question
is the same as to a similar question for four-vectors: one has to find
several quantities that under Lorentz transformations and translations in
flat space-time transform as components of a five-vector or of some other
five-tensor. Since one is talking about components, one has to specify the
basis in which they are evaluated. This is a simple matter if the definition
of the quantities one considers involves only scalars and components of
four-tensors in a Lorentz basis: since in either case $\nabla_{\mu} =
\partial_{\mu}$, the same should be true for the quantities defined, and
considering that in this basis $g_{\mu \nu} = \eta_{\mu \nu}$, one concludes
that the five-tensor components should correspond to a $P$-basis and
consequently should transform according to formulae (33).

It is apparent that this method of searching for five-tensors fails if
$\kappa = 0$, since in this case the transformation formulae do not enable
one to distinguish the components of a five-tensor from components of several
four-tensors. At $\kappa \neq 0$ the method works, but it does not allow one
to determine the precise value of $\kappa$. Indeed, if one has, say, five
quantities, $v^{A}$, that transform according to formulae (33) at a certain
value of $\kappa$, one can always construct five other quantities:
\begin{displaymath}
u^{\alpha} = v^{\alpha} \mbox{ and } u^{5} = \lambda v^{5},
\end{displaymath}
where $\lambda$ is an arbitrary nonzero constant, which will transform as
\begin{displaymath}
u'^{5} = u^{5} - (\lambda \kappa) a_{\alpha}
\Lambda^{\alpha}_{\, \beta} \, u^{\beta} \; \; \mbox{ and } \; \;
u'^{\alpha} = \Lambda^{\alpha}_{\, \beta} u^{\beta}.
\end{displaymath}
So as far as transformation laws are concerned, this quintuple may correspond
to a five-vector at any nonzero $\kappa$.

In the following I will suppose that $\kappa \neq 0$. In this case it is
convenient to slightly modify the definitions of the $O$- and $P$-bases by
taking that in both cases the fifth basis vector is normalized to $|\kappa|$
rather than to unity. In other words, it will be taken that ${\bf e}_{5} =
{\bf p}_{5} = \kappa {\bf n}$, where ${\bf n}$ is one of the two normalized
directional vectors of $V_{4}$. Such a change in the definitions results in
that the constant $\kappa$ disappears from formulae (26) and (33) and the
latter acquire a simpler form:
\begin{equation} \begin{array}{l}
{\bf p}_{\alpha}(x) = {\bf e}_{\alpha}(x) + x_{\alpha} {\bf e}_{5}(x) \\
{\bf p}_{5}(x) = {\bf e}_{5} (x).
\end{array} \end{equation}
and
\begin{flushright}
\hspace{4ex} \hfill $\left\{ \begin{array}{l} v'^{\alpha} =
\Lambda^{\alpha}_{\, \beta} v^{\beta} \\ v'^{5} = v^{5} - a_{\alpha}
\Lambda^{\alpha}_{\, \beta} \, v^{\beta} \end{array} \right.$ \hspace{4.5ex}
\hfill (35a) \end{flushright} and \begin{flushright}
\hspace{4ex} \hfill $\left\{ \begin{array}{l} w'_{\alpha} = w_{\beta}
(\Lambda^{-1})^{\beta}_{\, \alpha} + a_{\alpha} w_{5} \\ w'_{5} = w_{5}.
\end{array} \right.$ \hfill (35b) \end{flushright} \setcounter{equation}{35}

\vspace{2ex} \begin{flushleft}
\rm B. \it Covariant Lorentz coordinates and parameters \\
\hspace*{2.5ex} of Poincare transformations
\end{flushleft}
The simplest example of quantities that transform as components of
a nontrivial five-tensor are covariant Lorentz coordinates. Comparing
formula (44) of Appendix with formulae (35), we see that under Lorentz
transformations and translations the five quantities $x_{A}$, where
$x_{5} \equiv 1$, transform as components of a five-vector 1-form.
Consequently, if $\widetilde{\bf q}^{A}$ is the basis of five-vector
1-forms dual to the $P$-basis associated with the selected Lorentz
coordinate system, the 1-form $\widetilde{\bf x}$ constructed according
to the formula
\begin{equation}
\widetilde{\bf x} (x) \, \equiv \, x_{\alpha} \widetilde{\bf q}^{\alpha}(x)
+ \widetilde{\bf q}^{5}(x),
\end{equation}
will be the same no matter which system of Lorentz coordinates is
used.\footnote{As it has already been noted, for each Lorentz coordinate
system there exist {\em two} associated $O$-bases differing from each other
only in the overall sign of the basis five-vectors. By virtue of equations
(34), the same is true of the $P$-bases: there are two of them, and when
constructing the 1-form $\widetilde{\bf x}$ corresponding to the quintuple
$x_{A}$ one may use either of them. The 1-forms obtained with these two
bases will apparently differ in the sign. However, this ambiguity is of no
significance to us, since the results obtained below will be the same no
matter which of the two $P$-bases is selected.}

>From equations (34) one can easily obtain the formulae that relate the
basis $\widetilde{\bf q}^{A}$ to the basis of five-vector 1-forms
$\widetilde{\bf o}^{A}$ dual to the $O$-basis corresponding to the
same coordinates:
\begin{equation}
\widetilde{\bf q}^{\alpha}(x) = \widetilde{\bf o}^{\alpha}(x) \;
\mbox{ and } \; \widetilde{\bf q}^{5}(x) = \widetilde{\bf o}^{5}(x)
- x_{\alpha} \widetilde{\bf o}^{\alpha}(x).
\end{equation}
Substituting these relations into definition (36), one obtains the
following expression for the 1-form $\widetilde{\bf x}$ in the basis
$\widetilde{\bf o}^{A}$:
\begin{displaymath}
\widetilde{\bf x} (x) = x_{\alpha} \widetilde{\bf o}^{\alpha}(x)
+ \widetilde{\bf o}^{5}(x) - x_{\alpha} \widetilde{\bf o}^{\alpha}(x)
= \widetilde{\bf o}^{5}(x),
\end{displaymath}
from which one can clearly see that $\widetilde{\bf x}$ is indeed
independent of the choice of the coordinate system.

Let us also evaluate the covariant derivative of the field
$\widetilde{\bf x}$. Since in the $P$-basis all five-vector connection
coefficients are zero, one has
\begin{equation}
\nabla_{\mu} \widetilde{\bf x} = \partial_{\mu} x_{\alpha} \cdot
\widetilde{\bf q}^{\alpha} = \eta_{\mu \alpha} \widetilde{\bf q}^{\alpha}.
\end{equation}
The same result can be obtained in the $O$-basis, if one considers that in
this case the only nonzero connection coefficients are $G^{5}_{\; \alpha
\mu} = - \, \eta_{\alpha \mu}$, and so
\begin{displaymath}
\nabla_{\mu} \widetilde{\bf x} = \nabla_{\mu} \widetilde{\bf o}^{5} = - \,
G^{5}_{\; A \mu} \widetilde{\bf o}^{A} = - \, G^{5}_{\; \alpha \mu}
\widetilde{\bf o}^{\alpha} = \eta_{\mu \alpha} \widetilde{\bf o}^{\alpha},
\end{displaymath}
which on account of the first of equations (37), coincides with result (38).

Another example of purely geometric quantities that transform as components
of a nontrivial five-tensor are parameters of Poincare transformations.
When formulating the symmetry properties of flat space-time in section 4,
I have used Lorentz coordinates only as a tool for constructing the
equivalent fields. By itself, the replacement of a given set of fields
with an equivalent set, which is nothing but an active field transformation,
is an invariant procedure and can be considered without referring to any
coordinates. However, depending on how the latter are selected, a given
field transformation will correspond to different coordinate transformations.
Let us now find how the parameters of these coordinate transformations change
as one passes from one system of Lorentz coordinates to another.

The idea of the following calculation is very simple. One selects some set
of fields and a system of Lorentz coordinates, and by means of an arbitrary
Poincare transformation constructs the equivalent set of fields. One then
considers another system of Lorentz coordinates and determines the precise
Poincare transformation that one has to make in these new coordinates to
obtain the same set of equivalent fields. Finally, one expresses the
parameters of this second Poincare transformation in terms of the
parameters of the first one.

As a set of fields it is convenient to choose the covariant coordinates
associated with the selected Lorentz coordinate system $x^{\alpha}$, i.e.\
four scalar fields $\varphi_{(\alpha)}$ ($\alpha = 0, 1, 2, 3$) such that
\begin{displaymath}
\varphi_{(\alpha)}(Q) = \eta_{\alpha \beta} x^{\beta}(Q)
\end{displaymath}
at every point $Q$. Let us consider an arbitrary Poincare transformation
that corresponds to the coordinate transformation
\begin{equation}
x_{\alpha} \rightarrow y_{\alpha} =  x_{\beta} L^{\beta}_{\; \alpha}
+ b_{\alpha}.
\end{equation}
By virtue of the symmetry principle, the equivalent fields obtained by this
transformation are
\begin{displaymath}
\varphi^{\rm equiv}_{(\alpha)} = y_{\alpha} = x_{\beta} L^{\beta}_{\; \alpha}
+ b_{\alpha}.
\end{displaymath}
Let us now consider another system of Lorentz coordinates:
\begin{displaymath}
x'^{\alpha} = \Lambda^{\alpha}_{\; \beta} x^{\beta} + a^{\alpha}.
\end{displaymath}
In these coordinates the original fields acquire the form
\begin{displaymath}
\varphi_{(\alpha)} = (x'_{\beta} - a_{\beta}) \Lambda^{\beta}_{\; \alpha},
\end{displaymath}
and the equivalent fields are
\begin{displaymath}
\varphi^{\rm equiv}_{(\alpha)} = (x'_{\gamma} - a_{\gamma})
\Lambda^{\gamma}_{\; \beta} L^{\beta}_{\; \alpha} + b_{\alpha}.
\end{displaymath}
One should now present the right-hand side of the latter equation as
\begin{displaymath}
\varphi^{\rm equiv}_{(\alpha)} = (y'_{\beta} - a_{\beta})
\Lambda^{\beta}_{\; \alpha},
\end{displaymath}
where
\begin{displaymath}
y'_{\alpha} \equiv x'_{\beta} L'^{\beta}_{\; \alpha} + b'_{\alpha},
\end{displaymath}
and then express $L'^{\beta}_{\; \alpha}$ and $b'_{\alpha}$ in terms of
$L^{\beta}_{\; \alpha}$ and $b_{\alpha}$. Straightforward calculations give
\begin{equation} \begin{array}{l}
L'^{\alpha}_{\; \beta} = \Lambda^{\alpha}_{\; \sigma} L^{\sigma}_{\; \tau}
(\Lambda^{-1})^{\tau}_{\; \beta} \\ b'_{\beta} = b_{\tau}
(\Lambda^{-1})^{\tau}_{\, \beta} + a_{\beta} - a_{\rho} \Lambda^{\rho}_{\;
\sigma} L^{\sigma}_{\; \tau}(\Lambda^{-1})^{\tau}_{\; \beta},
\end{array} \end{equation}
which shows that the quantities ${\cal T}^{A}_{\; B}$ defined as
\begin{displaymath}
{\cal T}^{\alpha}_{\; \beta} = L^{\alpha}_{\, \beta},
 \; {\cal T}^{5}_{\; \beta} = b_{\beta}, \; {\cal T}^{\alpha}_{\; 5} = 0,
 \; \mbox{ and } \; {\cal T}^{5}_{\; 5} = 1,
\end{displaymath}
transform as components of a five-tensor of rank $(1,1)$.

It is also interesting to find the transformation formulae for the parameters
of infinitesimal Poicare transformations. In this case the matrix
$L^{\alpha}_{\; \beta}$ in equation (39) can be presented as
\begin{displaymath}
L^{\alpha}_{\; \beta} = \delta^{\alpha}_{\; \beta} +
{\scriptstyle \frac{1}{2}} ( \delta^{\alpha}_{\nu} \eta_{\beta \mu} -
 \delta^{\alpha}_{\mu} \eta_{\beta \nu} ) \, \omega^{\mu \nu},
\end{displaymath}
where $\omega^{\mu \nu} = - \omega^{\nu \mu}$, and both $\omega^{\mu \nu}$
and $b_{\alpha}$ are infinitesimals. From formulae (40) one obtains
\begin{displaymath}
\omega'^{\mu \nu} = \Lambda^{\mu}_{\, \alpha} \Lambda^{\nu}_{\, \beta}
\omega^{\alpha \beta} \; \; \mbox{ and } \; \;
b'^{\mu} = \Lambda^{\mu}_{\, \nu} (b^{\nu} -  a_{\alpha}
\Lambda^{\alpha}_{\, \beta} \, \omega^{\nu \beta}),
\end{displaymath}
which shows that the quantities ${\cal R}^{AB}$ defined as
\begin{displaymath}
{\cal R}^{\mu \nu} = \omega^{\mu \nu}, \; {\cal R}^{\mu 5} =
 - {\cal R}^{5 \mu} = b^{\mu}, \; \mbox{ and } \; {\cal R}^{55} = 0,
\end{displaymath}
transform as components of an antisymmetric five-tensor of rank $(2,0)$.

Further discussion of tensors ${\cal T}^{A}_{\; B}$ and ${\cal R}^{AB}$
will be made in part III.

\vspace{2ex} \begin{flushleft}
\rm C. \it Stress-energy and angular momentum tensors
\end{flushleft}
Let us now consider an example of physical quantities that transform as
components of a five-tensor: the canonical stress-energy and angular
momentum tensors, $\Theta^{\mu}_{\alpha}$ and $M^{\mu}_{\alpha \beta}$.

Let us begin by writing out the formulae that express the components of
these two tensors in one Lorentz coordinate system in terms of their
components in another Lorentz coordinate system. If the two coordinate
systems are related as in equation (30), then
\begin{equation} \left. \begin{array}{lcl}
\Theta'^{\mu}_{\; \alpha} & = & \Lambda^{\mu}_{\, \nu} \,
\Theta^{\nu}_{\beta} \, (\Lambda^{-1})^{\beta}_{\; \alpha}, \\
M'^{\mu}_{\; \alpha \beta} & = & x'_{\alpha} \Theta'^{\mu}_{\; \beta}
- x'_{\beta} \Theta'^{\mu}_{\; \alpha} + \Sigma'^{\mu}_{\; \alpha \beta} \\
& = & \Lambda^{\mu}_{\, \nu} M^{\nu}_{\sigma \tau}
(\Lambda^{-1})^{\sigma}_{\, \alpha} (\Lambda^{-1})^{\tau}_{\; \beta} \\
& & \hspace{6ex} + \; a_{\alpha} \Lambda^{\mu}_{\, \nu} \Theta^{\nu}_{\tau}
\, (\Lambda^{-1})^{\tau}_{\, \beta} \\ & & \hspace{12ex} -
\; a_{\beta} \Lambda^{\mu}_{\, \nu} \Theta^{\nu}_{\sigma} \,
(\Lambda^{-1})^{\sigma}_{\, \alpha},
\end{array} \right. \end{equation}
where $\Sigma^{\mu}_{\alpha \beta}$ is the spin angular momentum tensor.

With respect to their lower indices, $\Theta^{\mu}_{\alpha}$ and
$M^{\mu}_{\alpha \beta}$ are traditionally regarded as components
of four-tensors, and the fact that under space-time translations
$M^{\mu}_{\alpha \beta}$ acquires additional terms proportional to
$\Theta^{\mu}_{\alpha}$ is interpreted as a consequence of one actually
making a switch from one quantity---the angular momentum relative to the
point $x^{\mu} = 0$, to another quantity---the angular momentum relative
to the point $x'^{\mu} = 0$. Five-vectors enable one to give this fact a
different interpretation, which in several ways is more attractive.

One should notice that equations (41) coincide exactly with the
transformation formulae for components in the $P$-basis of a tensor---let
us denote it as $\cal M$---that has one (upper) four-vector index and
two (lower) five-vector indices and whose components are related to
$\Theta^{\mu}_{\alpha}$ and $M^{\mu}_{\alpha \beta}$ as follows:
\begin{equation} \begin{array}{l}
{\cal M}^{\mu}_{\alpha \beta} = M^{\mu}_{\alpha \beta}, \; \; {\cal M}
^{\mu}_{5 \alpha} \, = \Theta^{\mu}_{\alpha} \\ {\cal M}^{\mu}_{\alpha 5}
= -  \Theta^{\mu}_{\alpha}, \; \; {\cal M}^{\mu}_{5 5} = 0.
\end{array} \end{equation}
This coincidence means that $\Theta^{\mu}_{\alpha}$ and
$M^{\mu}_{\alpha \beta}$ can be regarded {\em as components of a
single five-tensor}. Since by definition $M^{\mu}_{\alpha \beta} =
- M^{\mu}_{\beta \alpha}$, this tensor is antisymmetric in its lower
(five-vector) indices.\footnote{More precisely, here $\cal M$ is regarded
as a dual of a four-vector 3-form whose values are covariant antisymmetric
five-tensors of rank 2.}

Such an interpretation of $\Theta^{\mu}_{\alpha}$ and
$M^{\mu}_{\alpha \beta}$ implies that there exists a single local physical
quantity: the stress--energy--angular momentum tensor $\cal M$. The belief
that there are many different angular momenta should now be regarded as
merely a wrong impression created by interpreting $\Theta^{\mu}_{\alpha}$
and $M^{\mu}_{\alpha \beta}$ as four-tensors: in reality, all these angular
momenta are components of $\cal M$ in different five-vector bases.

There is now no difficulty in defining the angular momentum density in
curved space-time. To see how this can be done, let us evaluate the
components of $\cal M$ in the $O$-basis. Using relations (37), one has
\begin{displaymath} \begin{array}{lll}
{\cal M} & = & (x_{\alpha} \Theta^{\mu}_{\beta} - x_{\beta}
\Theta^{\mu}_{\alpha} + \Sigma^{\mu}_{\alpha\beta}) \; \tilde{\bf q}^{\alpha}
\otimes \tilde{\bf q}^{\beta} \otimes {\bf E}_{\mu} \\ & & \hspace{8ex} + \;
(\Theta^{\mu}_{\beta}) \; \tilde{\bf q}^{5} \otimes \tilde{\bf q}^{\beta}
\otimes {\bf E}_{\mu} \\ & & \hspace{16ex} + \; (-\Theta^{\mu}_{\; \alpha})
\; \tilde{\bf q}^{\alpha} \otimes \tilde{\bf q}^{5} \otimes {\bf E}_{\mu} \\
& = & \Sigma^{\mu}_{\alpha \beta} \; \tilde{\bf o}^{\alpha} \otimes
\tilde{\bf o}^{\beta} \otimes {\bf E}_{\mu} \\ & & \hspace{8ex} + \;
(\Theta^{\mu}_{\beta}) \; \tilde{\bf o}^{5} \otimes \tilde{\bf o}^{\beta}
\otimes {\bf E}_{\mu} \\ & & \hspace{16ex} + \; (-\Theta^{\mu}_{\; \alpha})
\; \tilde{\bf o}^{\alpha} \otimes \tilde{\bf o}^{5} \otimes {\bf E}_{\mu}.
\end{array} \end{displaymath}
Thus, in the $O$-basis ${\cal M}^{\mu}_{\alpha \beta}$ coincide with the
components of the spin angular momentum tensor. In the case of flat
space-time one gives preference to the $P$-basis, since in it $\nabla_{\mu}
= \partial_{\mu}$, and, accordingly, the ${\cal M}^{\mu}_{\alpha \beta}$
components acquire additional terms proportional to covariant Lorentz
coordinates and to the components ${\cal M}^{\mu}_{\alpha 5}$ and
${\cal M}^{\mu}_{5 \beta}$. In the case of curved space-time, where a global
self-parallel basis does not exist, it is more convenient to use a regular
basis and have ${\cal M}^{\mu}_{\alpha \beta} = \Sigma^{\mu}_{\alpha \beta}$.

Let us now recall that canonical $\Theta^{\mu}_{\alpha}$ and
$M^{\mu}_{\alpha \beta}$ are defined as Noether currents corresponding to
Poincare transformations and as such satisfy the following ``conservation
laws'':
\begin{displaymath} \begin{array}{l}
\partial_{\mu} \Theta^{\mu}_{\alpha} = 0 \\ \partial_{\mu}
M^{\mu}_{\alpha \beta} = \eta_{\alpha \mu} \Theta^{\mu}_{\beta} -
\eta_{\beta \mu} \Theta^{\mu}_{\alpha} + \partial_{\mu}
\Sigma^{\mu}_{\alpha \beta} = 0.
\end{array} \end{displaymath}
One can now replace these two four-tensor equations with a single covariant
five-tensor equation:
\begin{equation}
{\cal M}^{\mu}_{\alpha \beta ; \mu} = 0,
\end{equation}
where it has been taken into account that in the $P$-basis all five-vector
connection coefficients are zero. It is interesting to see how equation
(43) works in the $O$-basis. One has
\begin{displaymath} \begin{array}{rcl}
{\cal M}^{\mu}_{5 \alpha ; \, \mu} & = & \partial_{\mu}
{\cal M}^{\mu}_{5 \alpha} - {\cal M}^{\mu}_{A \alpha} G^{A}_{\; 5 \mu} -
{\cal M}^{\mu}_{5A} G^{A}_{\; \alpha \mu} \\ & = & \partial_{\mu}
\Theta^{\mu}_{\alpha} - {\cal M}^{\mu}_{55} G^{5}_{\; \alpha \mu}
\; = \; \partial_{\mu} \Theta^{\mu}_{\alpha} \; = \; 0
\end{array} \end{displaymath}
and
\begin{displaymath} \begin{array}{rcl}
{\cal M}^{\mu}_{\alpha \beta ; \, \mu} & = & \partial_{\mu}
{\cal M}^{\mu}_{\alpha \beta} - {\cal M}^{\mu}_{A \beta}
G^{A}_{\; \alpha \mu} - {\cal M}^{\mu}_{\alpha A} G^{A}_{\; \beta \mu} \\
& = & \partial_{\mu} \Sigma^{\mu}_{\alpha \beta} - \Theta^{\mu}_{\beta}
G^{5}_{\; \alpha \mu} + \Theta^{\mu}_{\alpha} G^{5}_{\; \beta \mu} \\
& = & \partial_{\mu} \Sigma^{\mu}_{\alpha \beta} + \Theta^{\mu}_{\beta}
\eta_{\alpha \mu} - \Theta^{\mu}_{\alpha} \eta_{\beta \mu} \; = \; 0.
\end{array} \end{displaymath}
Thus, one obtains the same conservation laws for $\Theta^{\mu}_{\alpha}$
and $\Sigma^{\mu}_{\alpha \beta}$, only now the terms proportional to
$\Theta^{\mu}_{\alpha}$ in the second equation come from connection
coefficients.

\vspace{3ex} \begin{flushleft}
\bf Acknowledgements
\end{flushleft}
I would like to thank V. D. Laptev for supporting this work. I am grateful
to V. A. Kuzmin for his interest and to V. A. Rubakov for a very helpful
discussion and advice. I am indebted to A. M. Semikhatov of the Lebedev
Physical Institute for a very stimulating and pleasant discussion and to
S. F. Prokushkin of the same institute for consulting me on the Yang-Mills
theories of the de Sitter group. I would also like to thank L. A. Alania,
S. V. Aleshin, and A. A. Irmatov of the Mechanics and Mathematics Department
of the Moscow State University for their help and advice.

\vspace{3ex} \begin{flushleft} \bf
Appendix: Poincare transformation \\ \hspace{6ex} of covariant Lorentz
coordinates \\ \hspace{6ex} in the five-dimensional representation
\end{flushleft}
With any system of Lorentz coordinates, $x^{\alpha}$, in flat space-time
one can associate a system of covariant Lorentz coordinates defined as
$x_{\alpha} \equiv \eta_{\alpha \beta} x^{\beta}$, where $\eta_{\alpha \beta}
= {\rm diag}(+1, -1, -1, -1)$ is the Minkowski metric tensor. Under the
Poincare transformation
\begin{displaymath}
x'^{\alpha} = \Lambda^{\alpha}_{\; \beta} x^{\beta} + a^{\alpha},
\end{displaymath}
the covariant coordinates transform as
\begin{equation}
x'_{\alpha} = x_{\beta} (\Lambda^{-1})^{\beta}_{\, \alpha} + a_{\alpha},
\end{equation}
where $(\Lambda^{-1})^{\beta}_{\, \alpha}$ is the inverse of
$\Lambda^{\alpha}_{\, \beta}$ and $a_{\alpha} \equiv \eta_{\alpha \beta}
a^{\beta}$. Formally, one can present this inhomogeneous transformation
as a homogeneous transformation by introducing a {\em fifth} coordinate,
$x_{5}$, which is assigned a constant nonzero value, for example, $x_{5} =
1$.\footnote{A representation of this kind is used e.g. in the theory of
crystallographic groups.} Transformation (44) can then be presented as
\begin{displaymath}
x'_{A} = x_{B} L^{B}_{\, A},
\end{displaymath}
where $A$ and $B$ run 0, 1, 2, 3, and 5 and where
\begin{displaymath} \left \{ \begin{array}{l}
L^{5}_{\, 5} = 1, \; L^{\alpha}_{\, 5} = 0, \\ L^{5}_{\, \beta} = a_{\beta},
\; L^{\alpha}_{\, \beta} = (\Lambda^{-1})^{\alpha}_{\, \beta}.
\end{array} \right. \end{displaymath}
If $x_{5}$ is assigned some other nonzero value, say, $x_{5} = \kappa^{-1}$,
then instead of the latter formulae one will have
\begin{displaymath} \left \{ \begin{array}{l}
L^{5}_{\, 5} = 1, \; L^{\alpha}_{\, 5} = 0, \\ L^{5}_{\, \beta} = \kappa
a_{\beta}, \; L^{\alpha}_{\, \beta} = (\Lambda^{-1})^{\alpha}_{\, \beta}.
\end{array} \right. \end{displaymath}

\vspace{3ex} \begin{flushleft}
\bf References
\end{flushleft} \begin{enumerate}
 \item Th. Kaluza, {\it Sitzungsber. Preuss. Akad. Wiss. Berlin, Math.--Phys.
       K1.} (1921) 966; O. Klein, {\it Z. Phys.}, {\bf 46} (1927) 188.
 \item See e.g. K. S. Stelle and P. C. West, {\it Phys. Rev.} {\bf D21}
       (1980) 1466.
\end{enumerate}

\end{document}